# Direct Cryptographic Computation of the Cosmological Constant $\Omega_\Lambda$

Charles Kirkham Rhodes
Department of Physics, University of Illinois at Chicago,
Chicago, IL 60607-7059, USA

*Abstract*

A direct cryptographic computation of the Cosmological Constant $\Omega_\Lambda$ based solely on a physically anchored prime modulus $P_\alpha$ that stands in full agreement with observational data on $\Omega_\Lambda$ and $\Omega_m$ and the conclusion of a flat universe ($\Omega_\Lambda + \Omega_m = 1.0$) is demonstrated. The simplification derives from the fact that $\Omega_\Lambda$ defines the symmetry point of the cryptographic system.

## I. Introduction

Previous work [1] has shown how a cryptographic analysis [2] can relate the Higgs mass to the theorem of Quadratic Reciprocity [3], the founding statement of modern number theory first proved by Gauss. A subsequent study [4] showed that the Higgs state, a supersymmetric pair that defined the symmetry point of the cryptographic system, can be interpreted as the Cosmological Constants $\Omega_\Lambda$ and $\Omega_m$, with values that stand in full agreement with observation [5,6] and the condition for an exactly flat universe

$$\Omega_\Lambda + \Omega_m = 1.0 \quad , \tag{1}$$

as shown in Figs. (1) and (2). This latter outcome is legislated theoretically by the concept of supersymmetry [2], specifically, that the two residues $B_{Hh1}$ and $B_{Hh2}$ representing the Higgs states sum to modulus with the relation

$$B_{Hh1} + B_{Hh2} = P_\alpha^2 \tag{2}$$

in the quadratic extension field $\mathbb{F}_{P_\alpha^2}$. Related analyses [2,7,8] demonstrated the ability to provide both a theoretical basis and a <u>predicted magnitude</u> for the observed value [9-12] of the fine structure constant α that expresses a level of agreement with measured data of ~250 ppt, as quantitatively summarized in Fig. (3), using the identical cryptographic approach. An interesting aspect of the data presented in Fig. (3) is the discrepancy [8] between the direct determination of α with Rb recoils [10] and the corresponding indirect measurement of α that involves the experimental value of the electron g-2 combined with a tenth order QED analysis [11]. This difference points to the failure of the conventional QED picture at sufficiently high precision [8].

Overall, these results [1,2,4,7,8] culminated in the ability to show that $\Omega_\Lambda$, $\Omega_m$, and α are precisely related by a physically motivated algorithm and that the entire ($\Omega_\Lambda$, $\Omega_m$, α) triad is ultimately and perforce connected to Quadratic Reciprocity [3], hence also, the symmetry point of the cryptographic system. Thereby, basic physical entities align with a comparably fundamental mathematical statement. Physically, this means that (1) $\alpha$, $\Omega_\Lambda$, and $\Omega_m$ are rigorously connected, (2) <u>do not constitute independent quantities</u>, and (3) the calculation of any member of the ($\Omega_\Lambda$, $\Omega_m$, α) triple is likewise an equivalent calculation of the other two parameters[4]. These interlocking associations are, of course, mirrored precisely in the corresponding mathematical relations. Ultimately, these results construct a coherent synthesis [1,2,4,7,8], in full conformance with observational data, that quantitatively and mutually relates the six physically intrinsic universal parameters α, G, h, c, $\Omega_\Lambda$, and $\Omega_m$.

The organizing principle underlying these findings stems from the identification of an <u>observationally grounded</u> [1,2,4,7,8] modulus $P_\alpha$ and the subsequent construction of a modular counting system that correctly represents the physical entities. Equivalently, the theoretical picture transforms the physical analysis into a physically and mathematically doubly anchored cryptographic code-breaking problem in which the modulus $P_\alpha$ defines the counting rule.

Importantly, the computational system used is established a priori by measured data, contains no free or adjustable parameters, and hence, cannot utilize a fitting procedure a posteriori; accordingly, the numerical results shown in Figs. (1) to (3) stand or fall without a safety net.

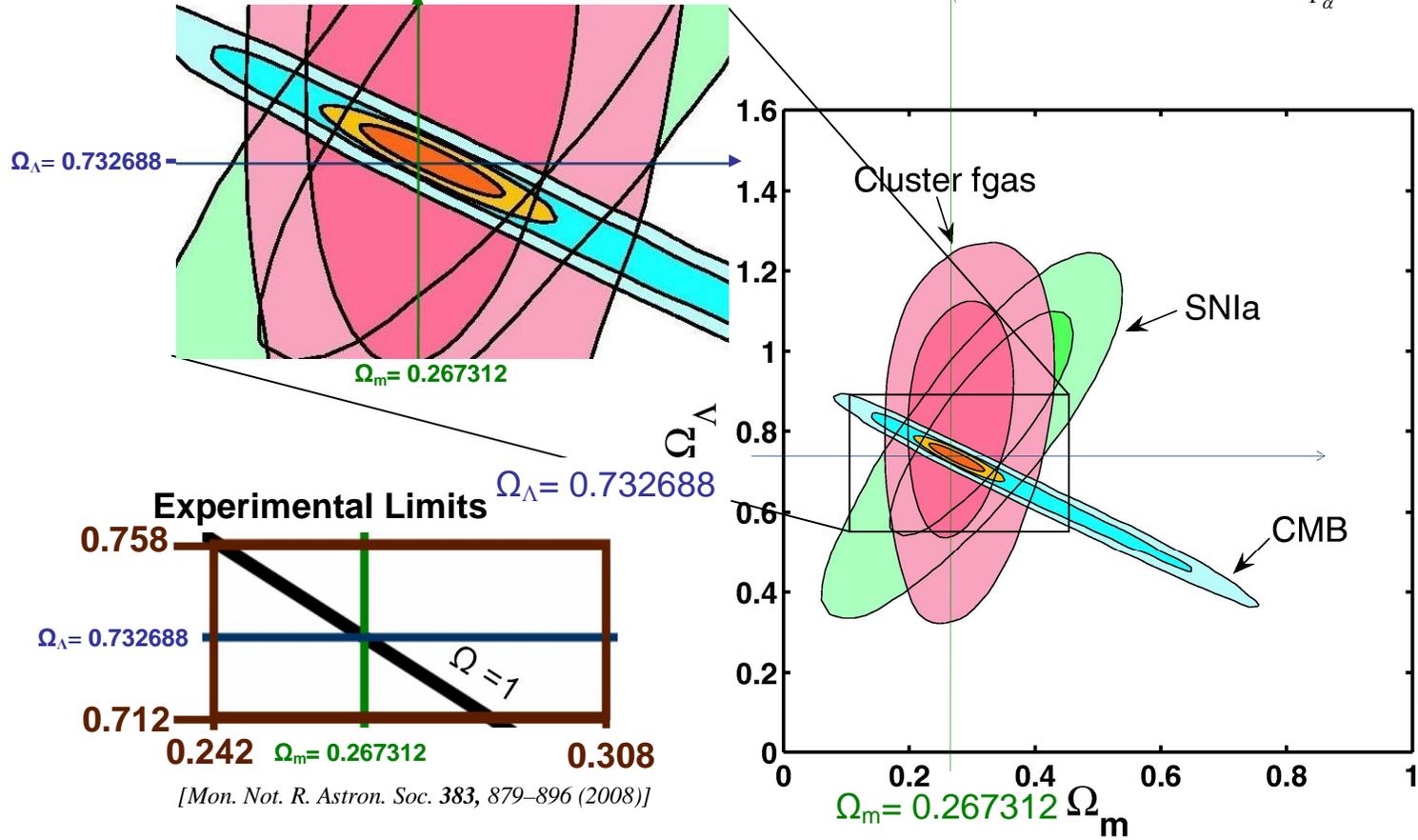

Fig. (1): Comparison of the computed values of $\Omega_\Lambda$ and $\Omega_m$ from the Super Higgs Congruence in $\mathbb{F}_{P_\alpha^2}$ with the assembly of correlated data restricting the ranges of $\Omega_\Lambda$ and $\Omega_m$. The concept of supersymmetry legislates the condition of $B_{Hh1}+B_{Hh2}= P_\alpha^2$, a statement equivalent to perfect flatness given by $\Omega_\Lambda + \Omega_m=1.0$. The inset shown at the upper left details the central zone illustrating the agreement between the calculated and experimental values. The theoretical values of $\Omega_\Lambda$ and $\Omega_m$ are compared directly with the experimental limits at 68% confidence ($0.712 < \Omega_\Lambda < 0.758$, $0.242 < \Omega_m < 0.308$) in the box placed in the lower left panel. The flat universe $\Omega =1$ contour is shown for reference. The figure is adopted from Ref. [5] and used with permission.

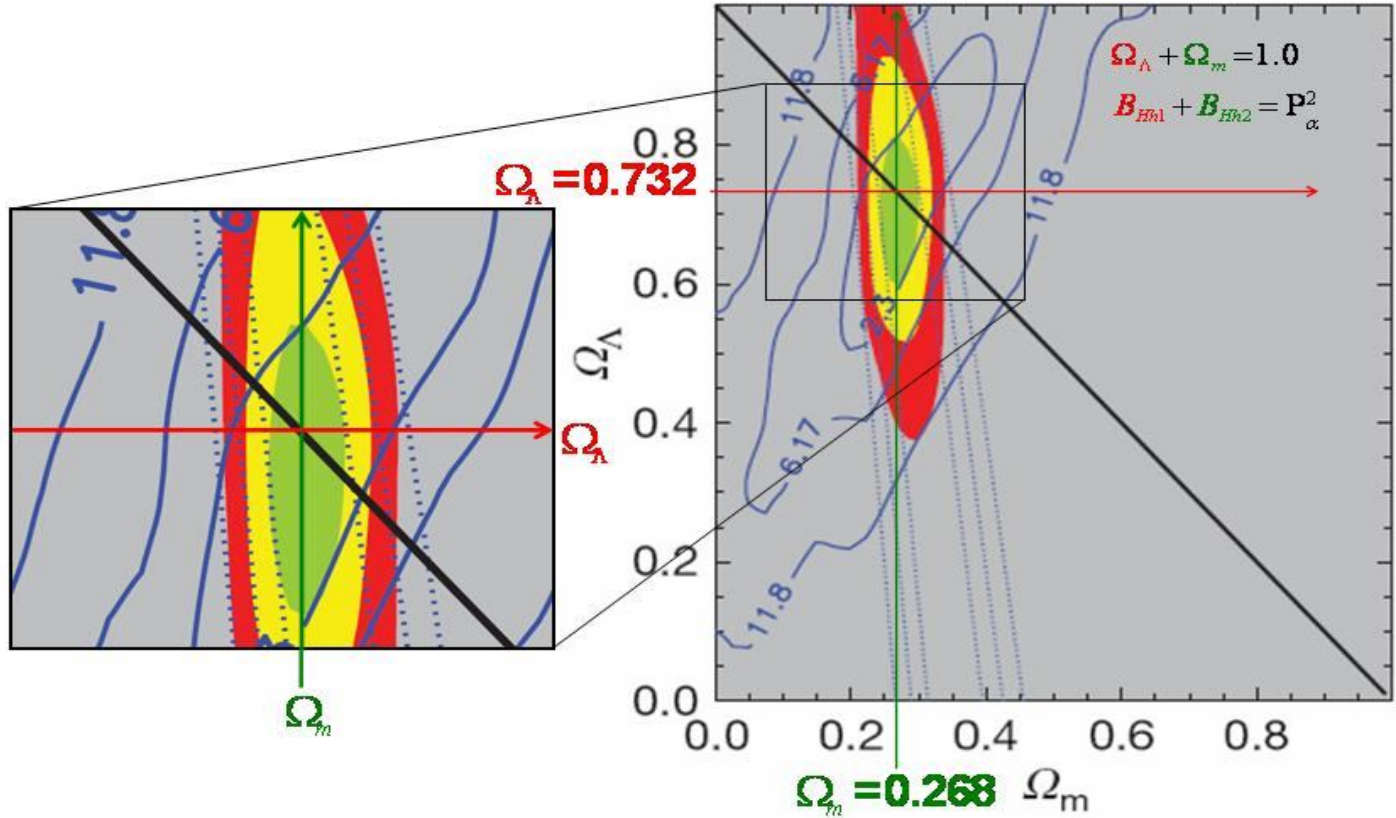

Fig. (2): Comparison of the computed values of $\Omega_\Lambda$ and $\Omega_m$ from the Super Higgs Congruence in $\mathbb{F}_{P_\alpha^2}$ with the assembly of correlated data restricting the ranges of $\Omega_\Lambda$ and $\Omega_m$ derived from a geometric measure based on bound galactic pairs [6]. Perfect agreement of the computed values with the data is manifest. The concept of supersymmetry legislates the condition of $B_{Hh1} + B_{Hh2} = P_\alpha^2$, a statement equivalent to perfect flatness given by $\Omega_\Lambda + \Omega_m = 1.0$, the condition specified by the black diagonal line. The figure is adapted from Ref. [6] and used with permission.

# Determination of the Fine Structure Constant
## Status November 2012

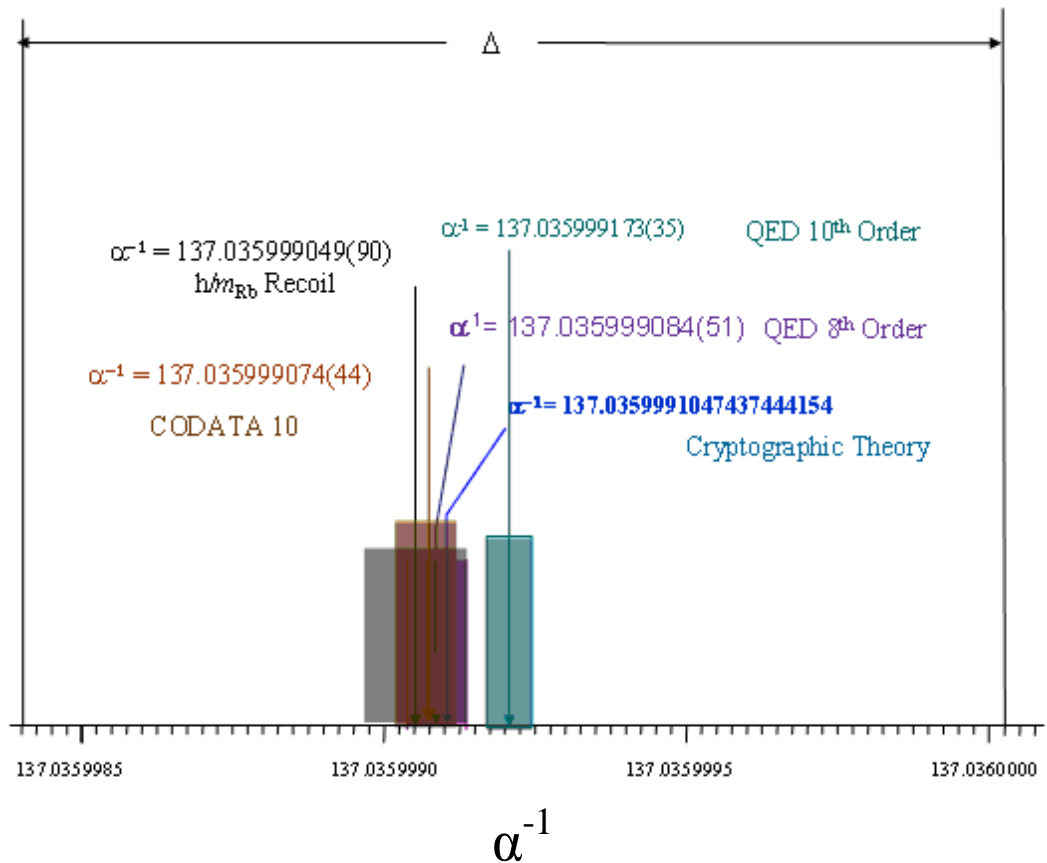

Fig. (3) Comparison of five values of the inverse of the fine structure constant $\alpha^{-1}$ obtained by various methods that presents the status as of November 2012. The direct non-QED-dependent value determined in 2011 from $h/m_{Rb}$ recoils [10,12] gives $\alpha^{-1}$=137.035999049(90); the value derived from a measurement [9] of the electron g-2 with eighth order QED analysis yields $\alpha^{-1}$=137.035999084(51) and with tenth order QED analysis [11] gives the revised value of $\alpha^{-1}$=137.035999173(35); the CODATA 2010 assessment [12] is stated as $\alpha^{-1}$=137.035999074(44); and the cryptographic [2] theoretical prediction (2003) has the value, to 22 digits, $\alpha^{-1}$= 137.0359991047437444154. The tenth order QED-dependent value [11] stands in disagreement with both the direct $h/m_{Rb}$ finding [10] and the eighth order QED-dependent result [9] based on the electron g-2 measurement. The cryptographic value is in agreement with both the non-QED-dependent $h/m_{Rb}$ recoil value and the CODATA 2010 assessment. Notably, the determination of $\alpha$ using the tenth order QED analysis [11] stands apart from all other values. The breadth $\Delta$ represents the level of uncertainty in $\alpha$ in 2003.

## II. Computations

### A. Higgs Congruence

The two fundamental congruences [1,2,4] governing the Higgs residues $B_{Higgs}$ are expressed respectively as

$$B_{Higgs}^2 \equiv -1 (\bmod P_\alpha) \tag{3}$$

and

$$B_{Higgs}^2 \equiv -1 (\bmod P_\alpha^2) \tag{4}$$

in the prime $\mathbb{F}_{P_\alpha}$ and quadratic extension $\mathbb{F}_{P_\alpha^2}$ fields. The solutions in the extension field $\mathbb{F}_{P_\alpha^2}$ correspond [4] to the Cosmological Constants $\Omega_\Lambda$ and $\Omega_m$. In both Eqs.(3) and (4), the solutions represent the subgroup of order four, the symmetry point of the cryptographic system.

By inspection, it is clear that the solutions of the congruences given by Eqs.(3) and (4) are entirely determined by the physically and mathematically defined [1,2,4] prime modulus $P_\alpha$, since no other quantity is present in the relationships. We will initially consider the solution of Eq. (3) and extend the procedure to Eq.(4).

### B. Solution of $B_{Higgs}^2 \equiv -1 (\bmod P_\alpha)$

The prime $P_\alpha$ obeys the congruence [1,2,4,7]

$$P_\alpha = 1 (\bmod 4), \tag{5}$$

and accordingly, on the basis of theorem of Fermat [13], can be represented uniquely as a sum of two squares given by the general form [14,15]

$$a_1^2 + b_1^2 = P_\alpha \tag{6}$$

with $a_1, b_1 \in \mathbb{Z}$. It follows directly by the Fibonacci identity [13] that the parallel statement

$$a_2^2 + b_2^2 = P_\alpha^2, \tag{7}$$

with $a_2, b_2 \in \mathbb{Z}$ also holds.

We now consider the solution of Eq.(3) for Higgs residue $B_{Higgs}$. Cassels has shown [16] that, given a modulus $M$ composed only of primes $p \equiv 1 \pmod 4$, the following statements hold:

If we consider pairs $(x,y) \in \mathbb{Z}^2$ that satisfy the congruence

$$x \equiv B_{Higgs} \, y \pmod{M}, \tag{8}$$

then

$$x^2 + y^2 \equiv 0 \pmod{M} \tag{9}$$

for all $(x, y)$ satisfying Eq.(8). Indeed, there exists a specific pair

$$(x, y) = (a, b) \tag{10}$$

such that

$$a^2 + b^2 = M. \tag{11}$$

Hence, with the choice

$$M = P_\alpha, \tag{12}$$

we obtain

$$a_1 \equiv b_1 B_{Higgs} \pmod{P_\alpha} \tag{13}$$

with $a_1$ and $b_1$ both defined by and known from Eq.(6). A direct solution is given by

$$B_{Higgs} \equiv a_1 b_1^{-1} \pmod{P_\alpha} \tag{14}$$

with $b_1^{-1}$ designating the inverse of the integer $b_1$.

## C. Solution of $B_{Higgs}^2 \equiv -1 (\mod P_\alpha^2)$

The extension of this method to the solution of Eq.(4) is immediate; it yields by inspection

$$B_{Higgs} \equiv a_2 b_2^{-1} (\mod P_\alpha^2) \tag{15}$$

with $(a_2, b_2) \in \mathbb{Z}^2$ and correspondingly given explicitly by Eq.(7). It remains to compute the pair $(a_2, b_2)$ from $(a_1, b_1)$. Elementary application of the Fibonacci identity [13] yields the results

$$a_2 = 2a_1 b_1 \tag{16}$$

and

$$b_2 = a_1^2 - b_1^2. \tag{17}$$

Since an earlier study [4] demonstrated that the Higgs states defined by Eq.(4) can be identified with the cosmological constants $\Omega_\Lambda$ and $\Omega_m$, we now have in Eq. (15) a direct computation of these quantities based purely on the representation of $P_\alpha$ given by Eq.(6).

## III. Numerical Results

The numerical findings are presented in Table I. Since $a_2$ and $b_2$ can play interchangeable roles, two solutions for $B_{Higgs}$ are produced that conform to Eq. (2); upon normalization [4] by $P_\alpha^2$, these solutions generate the corresponding values of $\Omega_\Lambda$ and $\Omega_m$. Perfect agreement with the earlier computations [4] is found, thereby yielding the identical values of $\Omega_\Lambda$ and $\Omega_m$ illustrated in Figs. (1) and (2). The precision of this agreement exceeds one part in $10^{121}$.

## IV. Conclusions

The cosmological constants $\Omega_\Lambda$ and $\Omega_m$ can be directly computed with high precision from the value of the physically and mathematically anchored [1,2,4,7,8] prime modulus $P_\alpha$ with an elementary algorithm. The values of both quantities stand in full agreement with observational data and the existence of a flat universe. The physical identification of these quantities is a supersymmetric particle pair representing the Higgs state.

**Acknowledgement:** This work was supported by Ultrabeam Technologies, LLC.

| Parameter | Integer | Prime Factors |
|---|---|---|
| $a_1$ | 5697822155169755616700610455226 | 2, 3, 7, 17, 1846817443, 432102431386199813 |
| $b_1$ | 25367172244133759839938099687625 | $5^3$, 37, 241, 45841, 4383991, 11324513171663 |
| $a_1^{-1}$ | 112942122406126866055968966494658727480100892492558403162752 | $2^7$, 3, 9293539, 15200459, 54341009, 56948712213581, 6727834295709363272907 |
| $b_1^{-1}$ | 59108095124890689158878934873057243535482651705375503040900525 | $5^2$, 19, 191, 14159, 22082245825229308231, 20837430518350292748876165542488 |
| $B_{Hl1}$ | 1464518162437886934175466847230415731421582430740004414836186 | 2, 3, 29, 37, 344353, 204901219, 14323448690905379, 2250856354248454454780404199 |
| $B_{Hl2}$ | 5295067887317048385691240239639344309098972470130691907921115 | 5, 13, 31, 109, 677, 25717, 165235181, 3158418359, 5645180551, 470014222057217424030 9 |
| $a_2$ | 289075272053265251135404154185359371403280058605374992763150 0 | $2^2$, 3, $5^3$, 7, 17, 37, 241, 45841, 4383991, 1846817443, 11324513171663, 432102431386199813 |
| $b_2$ | 6110282503516068946815233697321529377602389731063903833523949 | 41, 227104379, 16072168942091, 122381429412756089, 3336275564404308607 09 |
| $a_2^{-1}$ | 2171753445364098443948228612625334835003216764185679450634047829395975906 7946197052047834343087892920008323348479730103 48 | $2^2$, 92009, 5240000727176263007518368941, 11261308862899463941663603447445580853868860762059398 95643142417223698051777346389 28223 |
| $b_2^{-1}$ | 1945066274588647923994114234447531263644017764478762861205169531421671998 60549437434958135220620472647141351543951865423 31 | |
| $B_{Hh1}$ | 3347796095963539038288108836498780815207178149461268200903707119877006791 9259115346543183479220395326909020387514712622071 | $3^5$, 17, 23, 13221419329, 109323249654152734754556349619, 25581969464239851507538711579035419 93, 952908891213839759379015119512941741623 69 |
| $B_{Hh2}$ | 1221404260440614053082553050014296129165140090643353348094040656912288750 9222677465165575052088385852203865687752636182530 | 2, 5, 46663, 190633, 22176136014869, 43515013978215459227, 4137494096707187151247498420953 65017, 343895347592719957426696746853218547 605517 |
| $P_\alpha$ | 6759586049754935319866707086869760040520554900870696322757301 | $P_\alpha$ |
| $P_\alpha - 1$ | 6759586049754935319866707086869760040520554900870696322757300 | $2^2$, $3^2$, $5^2$, 7, 11, 13, 17, 19, 23, 29, 31, 37, 41, 43, 47, 53, 59, 61, 67, 71, 73, 79, 83, 89, 97, 101, 103, 107, 109, 113, 127, 131, 137, 139, 149, 151 |
| $P_\alpha + 1$ | 6759586049754935319866707086869760040520554900870696322757302 | 2, 8461, 45523, 83169760789807308284153, 10550476712756840929894132598 9 |
| $P_\alpha^2$ | 4569200356404153091370661886513076944372318240104621548997747776789295542 8481792811708758531308781179112886075267348804601 | $P_\alpha^2$ |

Table I: Computational results for the solutions of Eq. (14), giving $B_{Hl1}$ and $B_{Hl2}$, and Eq. (15), giving $B_{Hh1}$ and $B_{Hh2}$. The numerical figures stand in precise agreement with the earlier computations [4] of these quantities, confirming the analysis.